\documentclass[showpacs,preprintnumbers,amsmath,amssymb,axodraw,nofootinbib]{revtex4}

\usepackage{amsmath}
\usepackage{graphicx}
\usepackage{epsf}
\usepackage{psfrag}
\usepackage{epsfig}
\usepackage{graphics}
\usepackage[dvips]{color}

\begin{document}

\newcommand{\be}{\begin{equation}}
\newcommand{\ee}{\end{equation}}
\newcommand{\bq}{\begin{eqnarray}}
\newcommand{\eq}{\end{eqnarray}}

\title{Tree-level equivalence between a Lorentz-violating extension of QED  and its dual model in electron-electron scattering}

\date{\today}
\author{Giuliano R. Toniolo$^{(a)}$} \email[]{giulianortoniolo@hotmail.com}
\author{H. G. Fargnoli$^{(a)}$} \email[]{helvecio.fargnoli@dex.ufla.br}
\author{L. C. T. Brito$^{(a)}$} \email[]{lcbrito@dfi.ufla.br}
\author{A. P. Ba\^eta Scarpelli$^{(b),(c)}$} \email[]{scarpelli.apbs@dpf.gov.br}

\affiliation{(a) Universidade Federal de Lavras - Departamento de F\'{\i}sica \\
Caixa Postal 3037, 37.200-000, Lavras, Minas Gerais, Brazil}

\affiliation{(b)Setor T\'ecnico-Cient\'{\i}fico - Departamento de Pol\'{\i}cia Federal \\
Rua Hugo D'Antola, 95 - Lapa - S\~ao Paulo}

\affiliation{(c) Centro Federal de Educa\c{c}\~ao Tecnol\'ogica - MG \\
Avenida Amazonas, 7675 - 30510-000 - Nova Gameleira - Belo Horizonte
-MG - Brazil}

\begin{abstract}

\noindent
$S$-matrix amplitudes for the electron-electron scattering are calculated in order to verify the physical equivalence between two Lorentz-breaking dual models. We begin with an extended Quantum Electrodynamics which incorporates CPT-even Lorentz-violating kinetic and mass terms. Then, in a process of gauge embedding, its gauge-invariant dual model is obtained. The physical equivalence of the two models is established at tree-level in the electron-electron scattering and the unpolarized cross section is calculated up to second order in the Lorentz-violating parameter.

\end{abstract}


\maketitle

\section{Introduction}

In some situations, it is possible to establish relations between models which are essentially different but are equivalent in describing the physical behavior of a system. These are called dual models. This concept of duality is very useful, because there are some physical properties which are hidden in one model but are explicit in its dual theory. We refer to \cite{Polchinski} in order to exemplify this particularly interesting property of Quantum Field Theories. Different expansions for the same Hamiltonian in a quantum model can be written, as $H=H_0+ g H_1=H'_0+g' H'_1$, where $H_0$ and $H'_0$ allow simple known solutions. Besides, $H_0$ and $H'_0$ are expressed in a simple form in terms of the fields $\varphi$ and $\varphi'$, respectively. On the other hand, the relation between $\varphi$ and $\varphi'$ is complicated and nonlocal. Usually, the coupling constants obey a relation of the type $g \sim 1/g'$, so that $g'$ becomes small when $g$ is large and vice-versa. A very important fact is that if $g$ and $g'$ are not of the same order of magnitude, the description in terms of one of the fields will be appropriate for a perturbartive analysis. As a good example, the relation between electric and magnetic couplings is implemented by the dual mapping of a weakly-coupled theory in a strongly-coupled one.

We are interested in the kind of duality, investigated in the seminal work of Deser and Jackiw, between the three-dimensional spacetime self-dual and Maxwell-Chern-Simons models \cite{DJ}, which were discussed as a part of a wide class of models in \cite{Nieuw}. Since then, different techniques to attain the duality between models have been elaborated \cite{Polchinski,review}. Among the approaches to obtain physically equivalent models, we can cite the master action method \cite{Malacarne} and the gauge embedding technique \cite{Anacleto}. In the first approach, the so-called master action, roughly speaking, is written in terms of two vector fields. The dual models are then obtained by eliminating one of the fields from the action in favor of the other with the use of field equations. In the gauge embedding procedure (also called Noether dualization method), on the other hand, a gauge theory is obtained from a gauge-breaking model by the use of iterative embedding Noether counterterms, which vanish on mass shell. The Noether dualization method (NDM) is based on the idea of local lifting a global symmetry. This type of procedure is reminiscent of the earlier construction of component-field supergravity actions \cite{Nieuw2,Ferrara1,Ferrara2}. A particular interesting feature of such kind of dual models is that one can consider the non-invariant model as a gauge fixed version of a gauge theory. In other words, one model would reduce to the other under some gauge fixing conditions.

The gauging iterative Noether Dualization Method has been shown to be effective in establishing dualities between some models \cite{Ilha2}. This method provides a strong suggestion of duality, since it yields the expected result in the paradigmatic duality between the self-dual and Maxwell-Chern-Simons models in three dimensions. However, an intriguing result has been shown to be general when NDM is applied to Proca-like models \cite{scarp-epl}.  The gauge model obtained from the dualization algorithm, although sharing the physical spectrum with the original theory, acquires ghost modes. The gauge model obtained by means of gauge embedding encompasses the physical spectrum of the original Proca-like theory and, in addition, the spectrum of the corresponding massless model. However, these new modes appear with the wrong sign, characterizing ghosts, which may be dangerous for the model. In \cite{scarp-epl}, a relation between the propagators of dual models was obtained, which shed some light on this fact. Alternatives to avoid the emergence of ghosts in the process of dualization were studied \cite{Dalmazi1}, \cite{Dalmazi2}, \cite{Dalmazi3}, \cite{Dalmazi4}. In some cases, the price to be paid is the loss of locality \cite{Dalmazi1}. For a model with a spin-2 self-dual field in three spacetime dimensions, it was shown that the dual theory constructed with gauge embedding does not suffer with the presence of ghosts \cite{Dalmazi4}.

In some cases, such as the three-dimensional self-dual model, it is simple to see that these extra nonphysical modes are not harmful to the theory. This is because, in these cases, it is evident that the nonphysical particles have no dynamics or decouple from the rest of the model, since they do not contribute to the propagator saturated by conserved currents. Nevertheless, in some cases, it is not simple to check if the new modes spoil the gauge theory obtained with the process of dualization. Examples are the dualized Lorentz-violating models treated in the  papers \cite{Bota}, \cite{Petrov}, \cite{Fargnoli} and \cite{dual-LV}. It is not obvious that the ghosts and the physical particles decouple in these models. So, a deeper analysis is required.

In this paper, we will focus on the model of the reference \cite{Fargnoli}. The model is a modified QED, which incorporates two CPT-even Lorentz-breaking terms: a mass part of the type $-(1/2)m^2(g_{\mu \nu} - \beta b_\mu b_\nu)$ and the kinetic aether-like term of \cite{Carroll}, $- \frac \rho2 \left(b_\mu F^{\mu \nu}\right)^2$, in which $\beta$ and $\rho$ are dimensionless parameters and $b_\mu$ is a background vector. The model can be entirely accommodated in the Standard Model Extension (SME) \cite{kostelecky1,kostelecky2}, which provides a description of Lorentz and CPT violation in Quantum Field Theories, controlled by a set of coefficients whose small magnitudes are, in principle, fixed by experiments. The aether term is a particular case of the more general CPT-even Lorentz-violating kinetic part of SME. On the other hand, the Lorentz-violating mass term can be generated by spontaneous gauge symmetry breaking \cite{lv-mass1}, coming from the symmetric part of the second-rank background tensor which couples to the kinetic part of the Higgs field. Models with Lorentz-violating mass terms \cite{lv-mass2,lv-mass3,lv-mass4,lv-mass5} may present lots of interesting aspects, like superluminal modes or even instantaneous long-range interactions. The present model has been studied in many aspects in \cite{Fargnoli}, and the aforementioned properties were shown to appear for some values of the parameters $\beta$ and $\rho$. However, the dual gauge theory obtained in the process of dualization, as commented above, presents ghost modes whose role is still not clear. Since only the gauge sector has been studied, an analysis including the fermionic and interaction terms is missing.

In this paper, we reassess the model of \cite{Fargnoli} with the inclusion of the fermionic and the interaction sectors. The dualization by gauge embedding is carried out and the dual gauge invariant theory is obtained. In addition to the action achieved in the previous work, new interaction terms, which are nonminimal, are generated. We perform a practical calculation, the electron-electron scattering at tree-level, in order to check the decoupling of the nonphysical modes (calculations of scattering processes in Lorentz-violating models have been performed, for example, in the papers \cite{Manoel}). It is shown that nontrivial cancelations occur in the calculation with the dualized action, so that the two models yield identical results. Moreover, the unpolarized cross section was calculated up to second order in the Lorentz-violating parameter.

The paper is organized as follows. In section II, we describe and justify the model under analysis and, afterwards, we use the gauge embedding procedure to obtain its gauge invariant dual theory. In section III, the tree-level calculation of the electron-electron scattering is performed using the two models. We also obtain the unpolarized cross section up to second order in the Lorentz-violating parameter.  The concluding comments are in section IV.

\section{Description of the model and dualization}

In the present work, we consider the CPT-even Lorentz-breaking model of \cite{Fargnoli}, but now with  the vector field $A^{\mu}$ minimally coupled to a Dirac fermion. Thus, we have an extended QED model defined by the Lagrangian density
\begin{eqnarray}
\label{modelo}
\mathcal{L}^{(0)} & = & -\frac{1}{4}F_{\mu\nu}F^{\mu\nu}-\frac{\rho}{2}\left(b^{\mu}F_{\mu\nu}\right)^{2}+\frac{m^{2}}{2}A^{\mu}h_{\mu\nu}A^{\nu}+ \bar{\psi}\left[\gamma^{\mu}\left(i \partial_{\mu}+e A_{\mu}\right)-M\right]\psi,
\end{eqnarray}
where  $h_{\mu\nu} = g_{\mu\nu}-\beta b_{\mu}b_{\nu}$ and $b^\mu$ is a constant background four-vector. We should notice that the magnitude of $b^\mu$ is small compared to the other parameters of the theory. Here, $m$ and $M$ are the masses of the gauge field $A^{\mu}$ and the electron, respectively, while $\rho$ and $\beta$ are dimensionless parameters introduced simply to make the contributions from distinct Lorentz-violating terms, which appear in (\ref{modelo}), explicit. The kinetic Lorentz-violating term, which we call aether term \cite{Carroll}, is a particular version of the more general CPT-even part of the gauge sector of the Standard Model Extension \cite{kostelecky1, kostelecky2}, and can be radiatively induced \cite{aether} when nonminimal couplings to fermions \cite{nonminimal} are considered. On the other hand, the Lorentz-breaking mass term in the gauge sector may, for example,  be generated by spontaneous gauge symmetry breaking in a Lorentz-violating gauge-Higgs model \cite{lv-mass1}, emerging from the symmetric part of the second-rank background tensor which couples to the kinetic part of the Higgs field.

The gauge sector of this model was investigated in detail in \cite{Fargnoli} and it was shown that it incorporates very interesting features. For example, it presents physical massive poles which, depending on the choice of the coefficients $\rho$ and $\beta$, have their degrees of freedom changed. For this class of models, uncommon physical aspects can be accommodated for particular values of $\rho$ and $\beta$; for example, the presence of propagating superluminal modes.

We now proceed to the gauge embedding procedure. First, we calculate the variation of (\ref{modelo}) with respect to an infinitesimal change $\delta A^{\mu}$ in the gauge field:
\begin{eqnarray}
\delta\mathcal{L}^{(0)} & = & \left\{ \partial_{\beta}F^{\beta\mu}+\rho b^{\beta}b_{\alpha}\partial_{\beta}F^{\alpha\mu}-\rho b^{\mu}b_{\alpha}\partial_{\beta}F^{\alpha\beta}+m^{2}h^{\mu\alpha}A_{\alpha}+e\bar{\psi}\gamma^{\mu}\psi\right\} \delta A_{\mu} \nonumber \\
 & \equiv & J^{\mu}\delta A_{\mu}.
 \label{variat0}
\end{eqnarray}
It should be noticed that it is an off-shell method, since we have $J^{\mu}=0$ in the space of solutions. The current $J^{\mu}$ in (\ref{variat0}) can be used to construct a second  Lagrangian density,
\begin{equation}
\mathcal{L}^{(1)} = \mathcal{L}^{(0)}-B_{\mu}J^{\mu},
\label{L1}
\end{equation}
in which $B^{\mu}$ is an auxiliary vector field, chosen such that $\delta B_{\mu} = \delta A_{\mu}$. We calculate the variation of (\ref{L1}) with respect to $\delta A^\mu$ and get
\begin{equation}
\delta\mathcal{L}^{(1)} = -B_{\mu}\delta J^{\mu},
\label{deltaL1}
\end{equation}
with $\delta J_{\mu} = m^2h^{\mu\nu}\delta A_{\nu}$. Knowing the result (\ref{deltaL1}), we use a compensatory quadratic term  in the auxiliary field $B^{\mu}$ in order to build a gauge invariant Lagrangian density, given by
\begin{equation}
\mathcal{L}^{\left(2\right)}=\mathcal{L}^{\left(1\right)}+\frac{m^{2}}{2}B_{\mu}h^{\mu\nu}B_{\nu},
\end{equation}
in which it is simple to check that $\delta\mathcal{L}^{\left(2\right)}=0$. Finally, we calculate the variation of $\mathcal{L}^{\left(2\right)}$ with respect to $B^{\mu}$ to obtain
\begin{equation}\label{campoaux}
B_{\mu} = \frac{1}{m^2}L_{\mu\nu}J^{\nu},
\end{equation}
where we have defined the inverse of $h_{\mu \nu}$ as
\begin{equation}
L_{\mu\nu} = g_{\mu\nu} + \frac{\beta}{1-\beta b^2}b_{\mu}b_{\nu}.
\label{Lexp}
\end{equation}
Eq. (\ref{campoaux}) is used to write $\mathcal{L}^{\left(2\right)}$ in terms of the field $A^{\mu}$. The resulting dual gauge invariant Lagrangian associated with the original model (\ref{modelo}) reads
\begin{eqnarray}\label{lagdualfinal}
\mathcal{L}_{D}&\equiv&\mathcal{L}^{(2)}\nonumber\\&=&\frac{1}{4}F_{\mu\nu}F^{\mu\nu}+\frac{\rho}{2}\left(b^{\mu}F_{\mu\nu}\right)^{2}
-\frac{1}{2\alpha}\left(\partial_{\mu}A^{\mu}\right)^2
-\frac{1}{2m^{2}}\left(\partial_{\beta}F^{\beta\mu}\right)\left(\partial_{\sigma}F^{\sigma\nu}\right)L_{\mu\nu}\nonumber\\
&+&\frac{\rho}{m^{2}}b_{\alpha}\left[b^{\nu}\left(\partial_{\sigma}F^{\alpha\sigma}\right)\left(\partial_{\beta}F^{\beta\mu}\right)
-b^{\sigma}\left(\partial_{\sigma}F^{\alpha\nu}\right)\left(\partial_{\beta}F^{\beta\mu}\right)\right]L_{\mu\nu}\nonumber\\
&-&\frac{\rho^{2}}{2m^{2}}b_{\alpha}b_{\rho}\left[b^{\beta}b^{\sigma}\left(\partial_{\beta}F^{\rho\mu}\right)\left(\partial_{\sigma}F^{\alpha\nu}\right)
+b^{\mu}b^{\nu}\left(\partial_{\beta}F^{\rho\beta}\right)\left(\partial_{\sigma}F^{\alpha\sigma}\right)\right]L_{\mu\nu}\nonumber\\
&+&\bar{\psi}\left(i \gamma^\mu \partial_{\mu}+e \Gamma^\mu A_\mu -M\right)\psi
-\frac{e^{2}}{2m^{2}}\bar{\psi}\gamma^{\mu}\psi\bar{\psi}\gamma^{\nu}\psi L_{\mu\nu},
\label{duallagrang}
\end{eqnarray}
in which we have inserted a gauge fixing term and
\begin{eqnarray}
\Gamma^{\mu} &=& \frac{1}{m^2}\left\{\left[-L^{\mu\nu}\Box+L^{\alpha\nu}\partial^{\mu}\partial_{\alpha}\right]+\rho\left[-L^{\mu\nu}(b \cdot \partial)^2+L^{\alpha\nu}(b \cdot \partial)(b^{\mu}\partial_{\alpha}+b_{\alpha}\partial^{\mu})-b^{\mu}b_{\alpha}L^{\alpha\nu}\Box\right]\right\}\gamma_{\nu}.
\end{eqnarray}

By construction, the gauge embedding method gives rise to a gauge invariant Lagrangian, whereas the original model (\ref{modelo}) does not have this symmetry. It is believed that the non-invariant model can be considered as the gauge fixed version of a gauge theory. Note that now the Dirac fermions are nonminimally coupled  to the gauge field $A^{\mu}$. Besides, the dual Lagrangian has a contribution of a four-fermion nonrenormalizable vertex, which is similar to the result obtained in the duality between the self-dual and Maxwell-Chern-Simons models coupled to fermions \cite{Gomes:1997mf}.

\section{Electron-electron scattering}

We now proceed to perturbative calculations in order to check, in a practical calculation at tree level, the physical equivalence of the models. We first write the two photon propagators, which were obtained in \cite{Fargnoli}. From the quadratic terms in $A^{\mu}$, the propagators for the gauge field in momentum-space for the original (\ref{modelo}) and the dual (\ref{duallagrang}) models are given, respectively, by
\begin{eqnarray}
&&D^{\mu\nu}_{O}(k) =\frac{i}{A_1 H}\left\{-H \theta_{\mu \nu}
+ \frac{1}{m^2}\left[A_1 H +\beta \lambda^2(1+\rho b^2)A_1+(\rho+\beta)\lambda^2 m^2 \right] \omega_{\mu \nu} \right. + \nonumber \\
&&\left. +\left[(\rho + \beta) k^2 - \beta A_1 \right] \Lambda_{\mu \nu} - \lambda(\rho + \beta)\left( \Sigma_{\mu \nu} + \Sigma_{\nu \mu}\right)\right\},
\label{PropagOrig}
\end{eqnarray}
and
\begin{eqnarray}
&&D_{D}^{\mu\nu}(k)=\frac{i}{A_1 A_2 H}\left\{-m^2 H \theta_{\mu \nu} + \frac{1}{k^2}\left(-\alpha A_1 A_2 H+\frac{(1-\beta b^2)}{(1+\rho b^2)}\lambda^2m^2F \right) \omega_{\mu \nu} +
\right. \nonumber \\
&& \left.+ \frac{(1-\beta b^2)}{(1+\rho b^2)}m^2 F \Lambda_{\mu \nu} - \frac{(1-\beta b^2)}{(1+\rho b^2)}
\frac{\lambda m^2 F}{k^2} \left( \Sigma_{\mu \nu}+\Sigma_{\nu \mu} \right)\right\},
\label{PropagDual}
\end{eqnarray}
where
\begin{eqnarray}
A_1 &=& k^2-m^2+\rho \lambda^2, \nonumber\\
A_{2}&=&k^{2}+\rho\lambda^2, \nonumber\\
H &=& (1+\rho b^2)k^2 -(1-\beta b^2)m^2-\beta(1+\rho b^2)\lambda^2 \,\,\,\mbox{and} \nonumber\\
F&=&\rho A_{1}+\left(\rho+\beta\right)\frac{\left(1+\rho b^{2}\right)}{\left(1-\beta b^{2}\right)}k^{2}.
\end{eqnarray}

We have written the propagators in term of spin operators, being $\theta_{\mu \nu}=g_{\mu \nu}-\frac {k _\mu k _\nu}{k^2}$
and $\omega_{\mu \nu}=\frac {k _\mu k _\nu}{k^2}$ the transversal and the longitudinal operators, respectively. The operators $\Lambda _{\mu \nu}= b_\mu b_\nu$ and $\Sigma _{\mu \nu}=b_\mu k _\nu$ emerged from the inclusion of the external vector $b^\mu$ ($\lambda$ stands for $\Sigma _\mu \,\,^\mu=b_\mu k^\mu$). The Lorentz algebra of these operators is shown in Table 1:
\begin{center}
\begin{tabular}{|c|c|c|c|c|c|}
\hline
& $\theta _{\,\,\,\,\,\nu }^{\alpha }$ & $\omega _{\,\,\,\,\,\nu }^{\alpha }$ & $\Lambda _{\,\,\,\,\,\nu }^{\alpha }$ & $ \Sigma _{\,\,\,\,\,\nu }^{\alpha }$ & $\Sigma_\nu ^{\,\,\,\,\,\alpha}$ \\
\hline
$\theta _{\mu \alpha }$ & $\theta _{\mu \nu }$ & $0$ & $ \Lambda _{\mu \nu }-\frac{\lambda }{k^2}\Sigma _{\nu \mu }$ & $\Sigma_{\mu \nu }-\lambda \omega _{\mu \nu }$ & $0$ \\ \hline
$\omega _{\mu \alpha }$ & $0$ & $\omega_{\mu\nu}$ & $\frac{\lambda}{k^2}\Sigma _{\nu \mu }$ & $\lambda \omega _{\mu \nu }$ & $\Sigma_{\nu \mu}$ \\ \hline
$\Lambda _{\mu \alpha }$ & $\Lambda _{\mu \nu }-\frac{\lambda }{k^2} \Sigma_{\mu \nu }$ & $\frac{\lambda }{k^2}\Sigma_{\mu \nu }$ & $ b^{2}\Lambda _{\mu \nu }$ & $b^{2}\Sigma _{\mu \nu }$ & $\lambda \Lambda_{\mu \nu}$ \\ \hline
$\Sigma _{\mu \alpha }$ & $0$ & $\Sigma_{\mu \nu}$ & $\lambda \Lambda _{\mu \nu }$ & $\lambda \Sigma _{\mu \nu }$ & $k^2 \Lambda_{\mu \nu}$ \\ \hline
$\Sigma_{\alpha \mu}$ & $\Sigma_{\nu \mu} -\lambda \omega_{\mu \nu}$ & $\lambda \omega_{\mu \nu}$ & $b^2 \Sigma_{\nu \mu}$ & $b^2 k^2 \omega_{\mu \nu}$ & $\lambda \Sigma_{\nu \mu}$ \\ \hline
\end{tabular}
\vspace{2mm}

Table 1: Multiplicative table fulfilled by $\theta$, $\omega$, $\Lambda$
and $\Sigma$.
\end{center}

As carefully studied in \cite{Fargnoli}, the propagator $D_{D}^{\mu\nu}$ has, besides the physical poles of $D_O^{\mu \nu}$, new nonphysical ones. One way to proceed is to study the saturated propagator, which makes use of the current conservation to discard the nondynamical poles. Here, we intend to go further in a practical calculation of the $S$-matrix contribution at order $e^2$ for the electron-electron scattering, which is the main purpose  of this letter, and establish, for this process, the physical equivalence between the models.

\subsection{Calculation with the original model}

\begin{figure}[h]
\begin{center}
\includegraphics[scale=0.5]{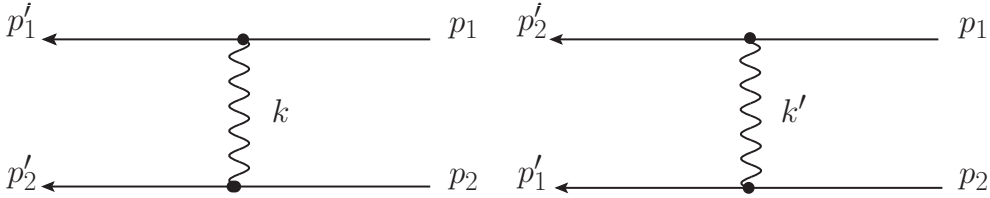}
\caption{Feynman diagrams for the electron-electron scattering at order $e^{2}$. In the original model, only these two  diagrams contribute. The gauge propagator is $D^{\mu\nu}_{O}$ and the vertex is the same as in ordinary QED. In the dual model, these diagrams must be accounted with the replacements $\gamma^{\mu}\rightarrow\Gamma^{\mu}$ and $D^{\mu\nu}_{O}\rightarrow D^{\mu\nu}_{D}$. The external lines represent on-shell Dirac electrons, where $p_1$ and $p_2$ are the momenta of the incoming particles, whereas $p'_1$ and $p'_2$ are the momenta of the outgoing particles. We have defined $k = p_1 - p^{\prime}_1$ and $k^{\prime} = p_1 - p^{\prime}_2$.}
\label{electron-3vertex}
\end{center}
\end{figure}

First, we consider the original model (\ref{modelo}). Since the model has only the usual Dirac fermion minimally coupled to the gauge field $A^{\mu}$, the two diagrams in figure \ref{electron-3vertex} contribute at tree-level with the same vertex as ordinary QED. However, in this case, with the $A^{\mu}$ propagator given by expression (\ref{PropagOrig}). The contribution to the $S$-matrix amplitude is given by $-(2\pi)^4\delta(p^{\prime}_{1}+p^{\prime}_{2}-p_{1}-p_{2})e^2 \tau_O$, where
\begin{eqnarray}
\tau_O&=&\bar{u}\left(p'_{1}\right)\gamma_{\mu}u\left(p_{1}\right)D^{\mu\nu}_{O}\left(k\right)\bar{u}\left(p'_{2}\right)\gamma_{\nu}u\left(p_{2}\right)\nonumber\\&-&\bar{u}\left(p'_{2}\right)\gamma_{\mu}u\left(p_{1}\right)D^{\mu\nu}_{O}\left(k'\right)\bar{u}\left(p'_{1}\right)\gamma_{\nu}u\left(p_{2}\right).
\label{amplit1Orig}
\end{eqnarray}
We have used $p_1$ and $p_2$ for external momentum of the free electrons in the initial states described by the spinors $u(p_{1})$ and $u(p_{2})$, and $p^{\prime}_1$ and $p^{\prime}_2$ for the free electrons in the final states $\bar{u}\left(p'_{1}\right)$ and $\bar{u}\left(p'_{1}\right)$. The expression (\ref{amplit1Orig}) can be obtained from the direct application of the LSZ reduction formula.

Since the fermions are on-shell, all terms from $D^{\mu\nu}_{O}$ which are dependent on the external momentum can be neglected in the calculation of the amplitude. Thus, we stay with
\begin{eqnarray}\label{amp_original}
\tau_O&=&\bar{u}\left(p'_{1}\right)\gamma_{\mu}u\left(p_{1}\right)\left[-\frac{1}{A_1(k)}\theta^{\mu\nu}+\frac{(\rho+\beta)k^2-\beta A_1(k)}{A_1(k)H(k)}\Lambda^{\mu \nu}\right]\bar{u}\left(p'_{2}\right)\gamma_{\nu}u\left(p_{2}\right)\nonumber\\
&-&\bar{u}\left(p'_{2}\right)\gamma_{\mu}u\left(p_{1}\right)
\left[-\frac{1}{A_1(k')}\theta^{\mu\nu}+\frac{(\rho+\beta)k'^2-\beta A_1(k')}{A_1(k')H(k')}\Lambda^{\mu \nu}\right]\bar{u}\left(p'_{1}\right)\gamma_{\nu}u\left(p_{2}\right).
\end{eqnarray}

\subsection{Calculation with the dual model}

\begin{figure}[h]
\center
\includegraphics[scale=0.5]{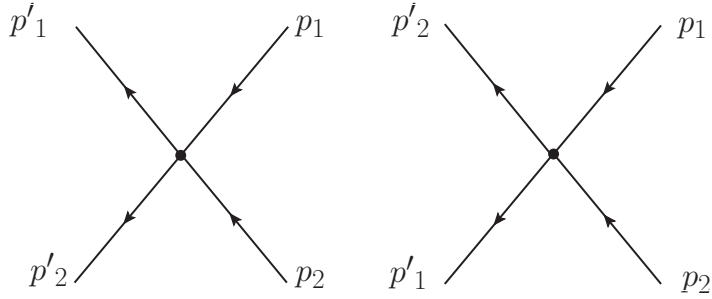}
\caption{Four-fermion vertex diagrams which contribute for the electron-electron scattering in the dual model. Again, $p_1$ and $p_2$ are the momenta of the incoming electrons, whereas $p'_1$ and $p'_2$ are the momenta of the outgoing electrons. These diagrams must be summed to the diagrams of figure \ref{electron-3vertex}.}
\label{electron-4vertex}
\end{figure}

We now proceed to the calculation of the tree-level electron-electron scattering by using the Feynman rules from the gauge model of (\ref{duallagrang}). Besides the two diagrams in figure \ref{electron-3vertex}, in the dual model we must take into account the diagrams of figure \ref{electron-4vertex}.
Now, for the calculation of the diagrams of figure \ref{electron-3vertex}, we must perform the replacements $\gamma^{\mu}\rightarrow\Gamma^{\mu}$ and $D^{\mu\nu}_{O}\rightarrow D^{\mu\nu}_{D}$. In momentum-space, we have
\be
\Gamma^\mu=\frac{1}{m^2}\gamma_\nu\left\{(k^2+\rho \lambda^2)\theta^{\mu \nu}+ \rho \lambda^2 \omega^{\mu \nu} + \frac{(\rho+\beta)}{(1-\beta b^2)} k^2
\Lambda^{\mu \nu} - \lambda \rho \Sigma^{\mu \nu}- \lambda \frac{(\rho+\beta)}{(1-\beta b^2)} \Sigma^{\nu \mu} \right\}
\ee
 and the calculation is greatly simplified if we note that
\begin{equation}
k_{\mu}\Gamma^{\mu}=0.
\label{relationGamma}
\end{equation}

By making these modifications in (\ref{amplit1Orig}), using the relation (\ref{relationGamma}) and the fact that fermions are on-shell, after a lengthy but straightforward algebra, we obtain
\begin{eqnarray}\label{tau1}
\tau_1&=&\bar{u}\left(p'_{1}\right)\gamma_{\mu}u\left(p_{1}\right)\left[\frac{\mathcal{Q}^{\mu\nu}
\left(k\right)}{m^2}\right]\bar{u}\left(p'_{2}\right)\gamma_{\nu}u\left(p_{2}\right)\nonumber\\
&-&\bar{u}\left(p'_{2}\right)\gamma_{\mu}u\left(p_{1}\right)\left[\frac{\mathcal{Q}^{\mu\nu}
\left(k'\right)}{m^2}\right]\bar{u}\left(p'_{1}\right)\gamma_{\nu}u\left(p_{2}\right),
\end{eqnarray}
in which
\begin{eqnarray}
&&\mathcal{Q}^{\mu\nu}\left(k\right)=-\frac{1}{A_1 A_2}\left\{ A_2^2 \theta^{\mu \nu}
+\frac{k^2}{(1-\beta b^2)^2}\left[(\rho+\beta)(\rho-\beta -2\rho \beta b^2) \lambda^2 \right. \right. \nonumber \\
&& \left. \left.+ (\rho + \beta)(2-\beta b^2 + \rho b^2)k^2
-(1+\rho b^2)(1- \beta b^2) \frac{F}{H k^2} (k^2-\beta \lambda^2)^2 \right]\Lambda^{\mu \nu}\right\}.
\label{Qexp}
\end{eqnarray}
In  addition, the contributions of the four-vertex diagrams of figure \ref{electron-4vertex} read
\begin{eqnarray}\label{tau2}
\tau_2 &=& \bar{u}\left(p'_1\right)\gamma_{\mu}u\left(p_1\right)\left[\frac{L^{\mu\nu}}{m^2}\right]\bar{u}\left(p'_2\right)\gamma_{\nu}u\left(p_2\right) \nonumber \\
&-&\bar{u}\left(p'_2\right)\gamma_{\mu}u\left(p_1\right)\left[\frac{L^{\mu\nu}}{m^2}\right]\bar{u}\left(p'_1\right)\gamma_{\nu}u\left(p_2\right).
\end{eqnarray}

Finally, putting together contributions (\ref{tau1}) and (\ref{tau2}), we obtain the total tree-level amplitude for the gauge-invariant theory:
\begin{eqnarray}
\tau_D &=& \bar{u}\left(p'_1\right)\gamma_{\mu}u\left(p_1\right)\frac{1}{m^2}\left[\mathcal{Q}^{\mu\nu}\left(k\right)+
L^{\mu\nu}\right]\bar{u}\left(p'_2\right)\gamma_{\nu}u\left(p_2\right) \nonumber \\
&-&\bar{u}\left(p'_2\right)\gamma_{\mu}u\left(p_1\right)\frac{1}{m^2}\left[\mathcal{Q}^{\mu\nu}\left(k'\right)+
L^{\mu\nu}\right]\bar{u}\left(p'_1\right)\gamma_{\nu}u\left(p_2\right).
\label{amplitDual}
\end{eqnarray}
Using expressions (\ref{Lexp}) and (\ref{Qexp}) for $L^{\mu\nu}$ and $\mathcal{Q}^{\mu\nu}$, respectively, and after a very lengthy algebra, we obtain
\begin{equation}\label{equivalencia}
\mathcal{Q}^{\mu\nu}+L^{\mu\nu} = \frac{m^2}{A_1}\left\{-\theta^{\mu\nu}+\frac{(\rho+\beta)k^2-\beta A_1}{H}\Lambda^{\mu \nu}\right\}.
\end{equation}
With this identity, we check that $\tau_O = \tau_D$ and the equivalence for this process is proved.

\subsection{The cross section}

To finish, we present the tree-level unpolarized cross section for the electron-electron scattering at second order in the Lorentz-violating background vector $b^{\mu}$. In the center-of-mass reference frame, the cross section is given by
\begin{equation}
\frac{d\sigma}{d\omega} = \frac{e^4M^4}{16\pi^2E_{CM}^2}|\tau|^2,
\end{equation}
where $\tau = \tau_O = \tau_D$ and $E_{CM}$ is the energy in the center of mass. Using the approximations $m/M << 1$ and $\mid b \mid^{2}<<1$, we obtain the following expansion
\begin{eqnarray}\label{secaochoque}
\left(\frac{d\sigma}{d\omega}\right)=\left(\frac{d\sigma}{d\omega}\right)_{QED} + \left(\frac{d\sigma}{d\omega}\right)_{Proca} + \rho\left(\frac{d\sigma}{d\omega}\right)^{\left(b^2\right)}_{LV}+\cdots,
\end{eqnarray}
in which the dots represent higher order terms in $b^\mu$. The first term in (\ref{secaochoque}) is just the well known result from ordinary QED. The second term is the contribution due to the Proca term, which reads\footnote{In these computations, we have used the FeynCalc Mathematica package.}
\begin{eqnarray}
\left(\frac{d\sigma}{d\omega}\right)_{Proca}=\frac{m^2\left\{\left(3E_{CM}^2-M^2\right)^2\left(1+\cos^2 \theta\right)^2+M^4\left(4+\sin^2 \theta\right)
-4 E_{CM}^4 \cos^4 \theta-8 E_{CM}^2M^2\right\}}
{64\pi^{2}E_{CM}^{2}\left(E_{CM}^{2}-M^2\right)^{3}\sin^6 \theta},
\end{eqnarray}
where $\theta$ is the scattering angle between the direction of the incident and the outgoing particles.  The last term of (\ref{secaochoque}) is the correction introduced by the Lorentz violation. Just to illustrate, we explicitly show the expressions for timelike and spacelike $b^\mu$. For spacelike $b^\mu$, we take $b^{\mu}=(\delta,0,0,0)$, such that the result reads
\begin{eqnarray}
\left(\frac{d\sigma}{d\omega}\right)_{LV}^{\left(b^{2}>0\right)}&=&\frac{\delta^{2}}{256\pi^{2}E_{CM}^{2}\left(E_{CM}^{2}-M^{2}\right)^{2}\sin^4 \theta}
\left\{3\left(E_{CM}^2-M^2\right)^2 \cos^4 \theta \right.\nonumber \\
&+&\left.\left(E_{CM}^2-M^2\right)\left(4E_{CM}^2-M^2\right)\cos^3 \theta
+\left(5E_{CM}^2-3M^2\right)\left(6E_{CM}^2+3M^2\right)\cos^2 \theta \right. \nonumber \\
&+&\left.\left(12 E_{CM}^4-3 E_{CM}^2M^2-M^4\right)\cos \theta + 15E_{CM}^4-15E_{CM}^2M^2+6M^4\right\}.
\end{eqnarray}
For the spacelike case, we use $b^{\mu} = (0,0,0,\delta)$ and the Lorentz-violating contribution reads
\begin{eqnarray}
\left(\frac{d\sigma}{d\omega}\right)_{LV}^{\left(b^{2}<0\right)}&=&
\frac{-\delta^{2}}{256\pi^{2}E_{CM}^{2}\left(E_{CM}^{2}-M^{2}\right)^{2}\sin^6 \theta}
\left\{E_{CM}^2\left(31 E_{CM}^2 + 40M^2\right)\cos^6 \theta\right.\nonumber\\
&+&\left. \left(E_{CM}^2-M^2\right)\left(4E_{CM}^2+M^2\right) \cos^5 \theta
+\left(199 E_{CM}^4 +149 E_{CM}^2M^2 + 2M^4\right) \cos^4 \theta \right. \nonumber \\
&+& \left.2\left(4 E_{CM}^4 + 5 M^4 \right) \cos^3 \theta + 3\left( 87 E_{CM}^4 + 35M^4\right) \cos^2 \theta \right. \nonumber \\
&-& \left. 3 \left( 4 E_{CM}^4 + 3M^4\right)\cos \theta + 3\left(7E_{CM}^4 + 4M^4\right)\right\}.
\end{eqnarray}

To carry out these calculations we have chosen the spatial part of $b^{\mu}$  in the same direction of the outgoing particle with momentum $p^{\prime}_{1}$. Note that at second order in $b^{\mu}$ only the aether term, $\frac{\rho}{2}\left(b^{\mu}F_{\mu\nu}\right)^{2}$, contributes to the cross section, since only the parameter $\rho$ appears in (\ref{secaochoque}). This is because in the expression for $\tau$, the coefficient of the transversal operator $\theta^{\mu \nu}=g^{\mu \nu}-\frac{k^\mu k^\nu}{k^2}$ does not depend on the $\beta$ parameter, which appears only in the coefficient of the operator $\Lambda^{\mu \nu}=b^\mu b^\nu$. Since we have calculated the unpolarized cross section, we expect that only the coefficient of the isotropic part contributes.

\section{Conclusion}

Dual models are constructed with the aim of having different descriptions of the same physical system but which are appropriate to be applied in distinct situations. Therefore, for some calculations, one of the models may furnish an obvious and simple result which is difficult to infer from the other one.
One of the artifacts of the dualization procedure by gauge embedding is the production, besides the original spectrum, of new nonphysical modes which, in some cases, may turn the model meaningless. These ghosts in some cases, such as in the famous duality between the three-dimensional self-dual and Maxwell-Chern-Simons models, are easily seen to have no dynamics. However, in most cases this is not an obvious issue, like in the Lorentz-breaking models studied in \cite{Fargnoli}.

In this paper, we showed that, sometimes, the physical equivalence of dual models is subtle. We carried out a practical calculation of a physical process, more precisely the cross section of the electron-electron scattering, using the dual model studied in detail in \cite{Fargnoli}. The equivalence of the models was shown at tree-level through nontrivial cancelations. For this, an essential role was played by the new fermionic couplings which emerged in the dualization process. Although these new modes apparently couple to the other sectors of the theory, these contributions are canceled out by other terms which come from new graphs due to this nonrenormalizable quartic vertex. Finally, the unpolarized cross section for this process was obtained. Besides the result from ordinary QED with a Proca term, new contributions were obtained up to second order in the background vector $b_\mu$. 

\vspace{1.0cm}

\noindent {\bf Acknowledgments}

\noindent A. P. B. S. acknowledges research grants from CNPq. Giuliano Toniolo thanks CAPES for the financial support.

\end{document}